\documentclass[epj]{svjour}
\usepackage{graphics}
\newcommand{\eq}{\begin{eqnarray}}
\newcommand{\en}{\end{eqnarray}}
\newcommand{\bea}{\begin{eqnarray}}
\newcommand{\eea}{\end{eqnarray}}

\newcommand{\ra}{\rangle}
\newcommand{\la}{\langle}

\begin{document}
\title{Electromagnetic properties of nucleons and hyperons\\
       in a Lorentz covariant quark model} 

\author{Amand Faessler\inst{1}, Thomas Gutsche\inst{1}, 
Barry R. Holstein\inst{2}, Valery E. Lyubovitskij\inst{1}, 
Diana Nicmorus\inst{1}, Kem Pumsa-ard\inst{1}}  
\offprints{}          % Insert a name or remove this line
\institute{\inst{1} Institut f\"ur Theoretische Physik,
Universit\"at T\"ubingen,\\ 
Auf der Morgenstelle 14, D-72076 T\"ubingen, Germany\\ 
\inst{2} Department of Physics--LGRT, University of
Massachusetts, Amherst, MA 01003 USA}
\date{Received: date / Revised version: date}
\abstract{
We calculate magnetic moments of nucleons and hyperons and
$N \to \Delta \gamma$ transition characteristics using
a manifestly Lorentz covariant chiral quark approach for the
study of baryons as bound states of constituent quarks dressed
by a cloud of pseudoscalar mesons.
} %end of abstract
\authorrunning{A. Faessler et al}  
\titlerunning{Electromagnetic properties of nucleons and hyperons 
in a Lorentz covariant quark model} 

\maketitle
\section{Introduction}
\label{intro}
The study of the magnetic moments of light baryons and of the
$N \to \Delta \gamma$ transition represents an old and important
problem in hadron physics. Many theoretical approaches -- lattice
QCD, QCD sum rules, Chiral Perturbation Theory (ChPT),
various quark and soliton methods, techniques based on the
solution of Bethe-Salpeter and Faddeev field equations,
etc. -- have been applied in order to calculate these quantities.

It should be stressed that analysis of the $N \to \Delta \gamma$
transition is of particular interest because it allows one to
probe the structure of both the nucleon and $\Delta(1232)$-isobar
and can help to shed light on their possible deformation. This
reaction represents a crucial test for the various theoretical
approaches. For example, naive quark models based on SU(6)
symmetry, which model the nucleon and its first resonance as a
spherically symmetric 3q-configurations, fail to correctly
describe the electric $G_{E2}$ and Coulomb $G_{C2}$ quadrupole
form factors, which vanish in such models in contradistinction
with experiment. A comprehensive analysis of the $N \to \Delta \gamma$ 
transition has been performed, e.g. in Refs.~\cite{Bjorken:1966ij}. 

There are a number of interesting problems which we
address in the present paper:

\begin{itemize}

\item [i)] if one believes that both valence and sea-quark effects
are important in the description of the electromagnetic properties
of light baryons, then how large is the contribution of the
meson-cloud;

\item [ii)] what is the physics required to correctly predict the $M1$
amplitude for the $N\to\Delta$ transition, which is considerably
underestimated in constituent quark models;

\item [iii)] what input is needed in order to explain the experimental
data for $E2/M1$ and $C2/M1$.

\end{itemize}

To possibly answer the above questions we use a Lo\-ren\-tz covariant
chiral quark model recently developed 
in Ref.~\cite{Faessler:2005gd,Faessler:2006ky}. 
The approach is based on a non-linear chirally symmetric
Lagrangian, which involves constituent quarks and the chiral
(pseudoscalar meson) fields as the effective degrees of
freedom. In a first step, this Lagrangian can be used to perform
a dressing of the constituent quarks by a cloud of light
pseudoscalar mesons and other heavy states using the calculational
technique of infrared dimensional regularization (IDR) of loop
diagrams. Then within a proper chiral expansion, we calculate the
dressed transition operators which are relevant for the
interaction of the quarks with external fields in the presence of
a virtual meson cloud. In a following step, these dressed
operators are used to calculate baryon matrix elements. Note,
that a simpler and more phenomenological quark
model which was based on the similar ideas of the dressing of the
constituent quarks by a meson cloud has been developed in
Refs.~\cite{PCQM}.

We proceed as follows. In Sec.~II, we
discuss basic notions of our approach. We derive 
a chiral Lagrangian motivated by baryon
ChPT~\cite{Becher:1999he}, and
formulate it in terms of quark and mesonic degrees of freedom.
Next, we use this Lagrangian to perform
a dressing of the operators of constituent quarks by 
a cloud of light pseudoscalar
mesons and by other heavy states. 
Then we discuss the calculation of matrix elements of dressed
quark operators between baryons states using a specific
constituent quark model~\cite{Ivanov:1996pz}-\cite{Faessler:2006ft}. 
In Sec.~III, we apply our approach to the study of magnetic
moments of light baryons and to the
properties of the $N\to\Delta\gamma$ transition.
In Sec.~IV we present a summary of our results.

\section{Approach}

The chiral quark Lagrangian ${\cal L}_{qU}$ (up to order $p^4$),
which dynamically generates the dressing of the constituent qu\-arks
by  mesonic degrees of freedom, consists of two primary pieces
${\cal L}_{q}$ and~${\cal L}_{U}$:
\eq\label{L_qU}
{\cal L}_{qU} &=& {\cal L}_{q} + {\cal L}_{U}\,, \ 
{\cal L}_q \, = \, {\cal L}^{(1)}_q + {\cal L}^{(2)}_q +
{\cal L}^{(3)}_q + {\cal L}^{(4)}_q + \cdots\,,\nonumber\\
{\cal L}_{U} &=&  {\cal L}_{U}^{(2)} + \cdots\,. 
\en
The superscript $(i)$ attached to ${\cal L}^{(i)}_{q(U)}$
denotes the low energy dimension of the Lagrangian. 
The detailed form of the chiral Lagrangian can be found
in Ref.~\cite{Faessler:2005gd,Faessler:2006ky}. Here for transparency 
we display only leading terms 
\eq\label{L_exp}
\hspace*{-.3cm}
{\cal L}_{U}^{(2)} &=&\frac{F^2}{4} \la{u_\mu u^\mu + \chi_+}\ra\,,  
{\cal L}^{(1)}_q =  \bar q [ i \, \slash\!\!\!\! D - m
+ \frac{1}{2} g  \slash\!\!\! u \, \gamma^5 ] q\,,  \nonumber \\
\en
where $q$ is the quark field, $u_\mu$ is the vielbein chiral field, 
the couplings $m$ and $g$ denote the quark mass and axial charge
in the chiral limit. The other notations are specified 
in~\cite{Faessler:2005gd,Faessler:2006ky}. 

Any bare quark operator (both one- and two-body) can be dressed by
a cloud of pseudoscalar mesons and heavy states in a
straightforward manner by use of the effective chirally-invariant
Lagrangian ${\cal L}_{qU}$. 
To calculate the electromagnetic transitions between baryons we project
the dressed electromagnetic quark operator $J_{\mu, \, {\rm em}}^{\rm dress}$  
between the corresponding baryon states.
The master formula is:
\eq\label{master_eq}
&&\la B(p^\prime) | \, J_{\mu, \, {\rm em}}^{\rm dress}(q)
\, | B(p) \ra \, = \, (2\pi)^4 \, \delta^4(p^\prime - p - q) 
\nonumber\\
&\times&\bar u_B(p^\prime) \biggl\{ \gamma_\mu \, F_1^B(q^2) \, + \,
\frac{i}{2 \, m_B} \, \sigma_{\mu\nu} q^\nu
\, F_2^B(q^2) \biggr\} u_B(p) \, \nonumber\\
&=& (2\pi)^4 \, \delta^4(p^\prime - p - q) [ M^V_\mu(q^2) + M^T_\mu(q^2) ] \, 
\en 
where 
\eq 
M^V_\mu(q^2) &=& \sum\limits_{q = u,d,s} f_D^q(q^2) \, 
\la B(p^\prime)|\,j_{\mu, q}^{\rm bare}(0)\,|B(p) \ra \,\\
M^T_\mu(q^2) &=& \sum\limits_{q = u,d,s}
 i \, \frac{q^\nu}{2 \, m_q} \, f_P^q(q^2) \,
\la B(p^\prime)| \, j_{\mu\nu, q}^{\rm bare}(0) \, |B(p) \ra \nonumber 
\en 
$B(p)$ and $u_B(p)$ are the baryon state and spinor, $m_B$ 
is the baryon mass. The explicit forms of
$f_D^q(q^2)$ and $f_P^q(q^2)$ are given in Ref.~\cite{Faessler:2005gd}. 
Here we focus on the diagonal $\frac{1}{2}^+
\to \frac{1}{2}^+$  transitions (the extension to the nondiagonal
transitions and transitions involving higher spin states like the
$\Delta(1232)$ isobar is straightforward). 
$F_1^B(q^2)$ and $F_2^B(q^2)$ are the Dirac and Pauli baryon
form factors. In the master equation (\ref{master_eq}) 
we express the matrix elements of the dressed quark operator 
in terms of the matrix elements of the  
bare operators. In our application we deal with the bare quark
operators for vector $j_{\mu, q}^{\rm bare}(0)$ and tensor
$j_{\mu\nu, q}^{\rm bare}(0)$ currents defined as
\eq\label{bare_operators}
j_{\mu, q}^{\rm bare}(0) \, = \, \bar q(0) \, \gamma_\mu \, q(0)\,,
\hspace*{.2cm} j_{\mu\nu, q}^{\rm bare}(0) \, = \, \bar q(0) \,
\sigma_{\mu\nu} \, q(0)\,. 
\en 
Equations~(\ref{master_eq})-(\ref{bare_operators}) contain 
our main result: we perform a model-independent
factorization of the effects of hadroni\-za\-ti\-on 
and confinement contained in
the matrix elements of the bare quark operators $j_{\mu, q}^{\rm bare}(0)$
and $j_{\mu\nu, q}^{\rm bare}(0)$ and the effects dictated by chiral
symmetry (or chiral dynamics) which are encoded in the relativistic form
factors $f_D^q(q^2)$ and $f_P^q(q^2)$. Due to this factorization
the calculation of $f_D^q(q^2)$ and $f_P^q(q^2)$, on one side,
and the matrix elements of $j_{\mu, q}^{\rm bare}(0)$ and
$j_{\mu\nu, q}^{\rm bare}(0)$, on the other side, can be done
independently. In particular, in a first step we derived
a model-independent formalism based on the ChPT Lagrangian,
which is formulated in terms of constituent quark degrees of freedom,
for the calculation of $f_D^q(q^2)$ and $f_P^q(q^2)$. 
Note, that the separate bare and the meson cloud contributions to the baryon 
form factors are defined as 
\eq 
M^{\rm bare}_\mu(q^2) &=& M^{V; 0}_\mu(q^2) \,, \\ 
M^{\rm cloud}_\mu(q^2)&=&  M^V_\mu(q^2) - M^{V; 0}_\mu(q^2) + M^T_\mu(q^2) 
\nonumber 
\en 
where 
\eq 
M^{V; 0}_\mu(q^2) &=& \sum\limits_{q = u,d,s} f_D^q(0) \, 
\la B(p^\prime)|\,j_{\mu, q}^{\rm bare}(0)\,|B(p) \ra 
\en 
and 
$f_D^q(0) = e_q$ is the quark charge. 

The calculation of the matrix elements of the bare
quark operators $j_{\mu, q}^{\rm bare}$ and $j_{\mu\nu, q}^{\rm bare}$  
can then be relegated to quark models based on specific
assumptions about had\-ro\-ni\-za\-ti\-on and confinement 
with taking into account of certain constraints dictated by 
Lo\-ren\-tz and gauge invariance and 
chiral symmetry~\cite{Faessler:2005gd,Faessler:2006ky}. 
Here we consistently employ the relativistic three-quark model 
(RQM)~\cite{Ivanov:1996pz}-\cite{Faessler:2006ft} to compute such
matrix elements. The RQM was previously
successfully applied for the study of properties of baryons
containing light and heavy
quarks~\cite{Ivanov:1996fj}-\cite{Faessler:2006ft}. The main
advantages of this approach are: Lorentz and gauge invariance, a
small number of parameters, and modelling of effects of strong
interactions at large ($\sim 1$ fm) distances. A preliminary
analysis of the electromagnetic properties of nucleons has been
performed in Ref.~\cite{Ivanov:1996pz} where the effects of
valence quarks have been consistently taken into account. 
Here we extend this analysis to the
case of hyperons as well as to the $N \to \Delta\gamma$
transitions and we include meson-cloud effects. The basic idea of RQM 
is to model the coupling of baryons to their valence quarks using 
the three-quark currents~\cite{Efimov:1987na} which are also extensively 
used in QCD sum rules~\cite{Ioffe:1982ce}. For the octet 
baryon states one can write two possible currents (vector 
$J_B^V$ and tensor $J_B^T$), while for the decuplet states 
only one current -- $J_D$. 
In particular, for the proton and $\Delta^+$-isobar the currents look as: 
\eq 
J_p^V &=& \varepsilon^{a_1a_2a_3} \gamma^\mu \gamma^5 d^{a_1} u^{a_2}
C \gamma_\mu u^{a_3} \nonumber\\
J_p^T &=& \varepsilon^{a_1a_2a_3} \sigma^{\mu\nu} \gamma^5 d^{a_1} u^{a_2}
C \sigma_{\mu\nu}  u^{a_3} \,\\
J_{\Delta^{+}}^\mu &=& \frac{1}{\sqrt{3}}
\varepsilon^{a_1a_2a_3} (d^{a_1} u^{a_2}
C \gamma^\mu u^{a_3}  +
2 u^{a_1} u^{a_2} C \gamma^\mu d^{a_3})
\,, \nonumber 
\en 
where $a_i$ are the color indices and $C$ is the charge conjugation matrix. 

\section{Physical applications}

In this section we consider the application of our technique to
the problem of magnetic moments of light baryons and the static
characteristics of the $N \to \Delta \gamma$ transition.  We
calculate the contributions of both valence and sea-quarks to
these quantities using the approach discussed above. In particular, 
we present results for magnetic moments of light baryons (Table 1)  
and properties of $N \to \Delta \gamma$ transition (Table 2):  
magnetic, electric and Coulombic form factors $G_{M1}$, 
$G_{E2}$ and $G_{C2}$, helicity amplitudes $A_{3/2}$ and $A_{1/2}$ at 
zero recoil, ratios of multipoles EMR = $E2/M1 = - G_{E2}/G_{M1}$ and 
CMR = $C2/M1 = - G_{C2}/G_{M1}$, transition dipole 
$\mu_{N\Delta}$ and quad\-ru\-po\-le $Q_{N\Delta}$ mo\-ment, 
$\Delta^+ \to p + \gamma$ decay width.  
In Tables 1 and 2 we show the contributions 
both of the valence quarks $(3q)$ and of the meson cloud and compare 
the total results with data~\cite{Yao:2006px}. 

As stressed above, for the octet states there exist two possible
choices for the three-quark current: vector and tensor. A
preliminary analysis (see also Ref.~\cite{Ivanov:1996pz,Faessler:2006ky}) 
showed that these two types of currents give practically the same 
(or at least very similar) results in the case of the static properties
of light baryons, {\it e.g.}, magnetic moments.  This result is
easily understood because the vector and tensor currents of the
baryon octet become degenerate in the nonrelativistic limit. Also,
the magnetic moments of light baryons are dominated by the
nonrelativistic contributions, with relativistic corrections being
of higher order and small.  This explains why the simple
nonrelativistic quark approaches work so well in the description
of the magnetic moments of light baryons. Therefore, in order to
distinguish between the two types of currents of the baryon octet
we need to examine quantities which are dominated by relativistic
effects.  Two such quantities are the well known ratios $E2/M1$
and $C2/M1$ of the multipole amplitudes characterizing the $N \to
\Delta \gamma$ transition. Here we find that the sole use of
vector and tensor currents gives {\it opposite} results for the
signs of these ratios. In particular, the use of the pure vector
current for the proton gives reasonable results for $E2/M1$ and
$C2/M1$ both with a {\it correct} (negative) sign, while the use
of the pure tensor current yields ratios with {\it wrong}
(positive) sign.  Therefore, the study of the ratios $E2/M1$ and
$C2/M1$ allows one to select the appropriate current for the
description of the bound-state structure of the baryon octet
(nucleons and hyperons). It is interesting to note that in the QCD
sum rule method~\cite{Ioffe:1982ce} dealing with current quarks
the vector current structure is also preferred. This choice
originally gave an explanation of the nucleon mass, while the use
of the tensor current yields a suppression of the nucleon mass due
to the ``bad'' chiral properties of this type of the three-quark
current. We would like to stress, however, that this preference of
the vector current for the description of the baryon octet in our
approach and in QCD sum rules is apparently just coincidental
because here we are dealing with constituent quarks instead of
current quarks. For the EMR and CMR ratios we present our 
predictions at zero recoil ($Q^2 = 0$) and at the finite value 
$Q^2 = 0.06$ GeV$^2$ (recently the A1 Collaboration at
Mainz~\cite{Stave:2006ea} measured these quantities at this
kinematic point). Our predictions are in good
agreement with the experimental data of the LEGS Collaboration
at Brookhaven~\cite{Blanpied:2001ae} and of the GDH, A1 and A2
Collaborations at Mainz~\cite{Stave:2006ea,Ahrens:2004pf}. 

\vspace*{1cm}

\noindent
{\bf Table 1.} Magnetic moments of light baryons. 

\begin{center}
\def\arraystretch{1.75}
\begin{tabular}{ccccc}
\hline

& Bare & Meson & Total & Data~\cite{Yao:2006px}\\ 
& (3q) & cloud &       & \\
\hline 
$\mu_p$ &\, 2.614 \,&\, 0.179 \,&\, 2.793 \,&\, 2.793 \\
$\mu_n$ & -1.634 & -0.279 & -1.913 & -1.913 \\
$\mu_{\Lambda}$ & -0.579 & -0.034 & -0.613 &-0.613 $\pm$ 0.004 \\
$\mu_{\Sigma^+}$ & 2.423 & 0.148 & 2.571 & 2.458 $\pm$ 0.010\\
$\mu_{\Sigma^-}$ & -0.960 & -0.223 & -1.183 & -1.160 $\pm$ 0.025\\
$\mu_{\Xi^0}$ & -1.303 & -0.082 & -1.385 &-1.250 $\pm$ 0.014 \\
$\mu_{\Xi^-}$ & -0.567 & 0.012 & -0.555 & -0.651 $\pm$ 0.003 \\
$|\mu_{\Sigma^0 \Lambda}|$ & 1.372 & 0.245 & 1.617 & 1.61 $\pm$ 0.08 \\
$\mu_{N \Delta}$ & 2.984 & 0.354 & 3.338 & 3.642 $\pm$ 0.019 \\
\hline
\end{tabular}
\end{center}

\vspace*{1cm}

\noindent 
{\bf Table 2.} Results for the
N $\to \Delta\gamma$ transition. Notations: \\ 
EMR and CMR in \% (subscripts 0 and 0.06 
mean the values of $Q^2 = 0$ and 0.06 GeV$^2$), 
$A_{i}$ in $10^{-3}\ $ GeV$^{-1/2}$, 
$Q_{N\Delta}$ in fm$^2$, 
$\Gamma_{\Delta \to N \gamma}$ in MeV. 

\begin{center}
\def\arraystretch{1.6}
\begin{tabular}{lcccc}
\hline
 & Bare  & Meson & \,\,Total \,\, &
Data~\cite{Yao:2006px}\\ 
 & (3q)  & cloud &                & \\ 
\hline 
EMR$_0$ & -3.41 & 0.31 & -3.10 & -2.5 $\pm$ 0.5\\
EMR$_{0.06}$ & -3.34 & 0.33 & -3.01
& -2.28 $\pm$ 0.29 \\
CMR$_{0}$ & -3.95 & 0.26 & -3.69 & \\
CMR$_{0.06}$ & -5.13 & 0.35 & -4.78
& -4.81 $\pm$ 0.27 \\
$A_{1/2}$ 
& -110.0 & -14.3 & -124.3 & -135 $\pm$ 6\\
$A_{3/2}$ 
& -219.4 & -25.3 & -244.7 & -250 $\pm$ 8\\
$G_{E2}$ & 0.125 &  0.002 & 0.127    & 0.137 $\pm$ 0.012 \\
$G_{M1}$ & 3.655 & 0.434 & 4.089 & 4.460 $\pm$ 0.023 \\
$G_{C2}$ & 0.144 & 0.007 & 0.151 & \\
$Q_{N\Delta}$ &  -0.098 & -0.001 & -0.099 &
-0.108 $\pm$ 0.009 \\
$\mu_{N\Delta}$ &  2.984 & 0.354 & 3.338
& 3.642 $\pm$ 0.019   \\
$\Gamma_{\Delta \to N \gamma}$  & 0.49  & 0.12 & 0.61
& 0.58 - 0.67 \\
\hline
\end{tabular}
\end{center}

\section{Summary}

In this paper we have calculated the magnetic moments of light
baryons as well as the $N \to \Delta \gamma$ transition
properties using a manifestly Lorentz covariant chiral quark
approach to the study of baryons as bound states of constituent
quarks dressed by a cloud of pseudoscalar mesons. Our main results
are:

- The contribution of the meson cloud to the static properties of
light baryons is up to 20\%, which is consistent with the
perturbative nature of their contribution and, together with the
relativistic corrections, helps to explain how the ~30\% shortfall
in the SU(6) prediction is ameliorated;

- We get a reasonable description for the dipole magnetic moment
$\mu_{N\Delta}$ due to the enhancement of the valence quark 
contribution~\cite{Faessler:2006ky}; 

- The multipole ratios EMR and CMR are sensitive to the choice of the
  proton current: vector $J_p^V$ or tensor $J_p^T$. 
  The use of a pure vector current $J_p^V$ gives a reasonable
  description of the data. The pure tensor current $J_p^T$ gives results
  for EMR and CMR with the wrong (positive) sign.  However, a small
  admixture of the tensor current is possible, and forthcoming
  experiments can give a strong restriction on the mixing parameter of
  such currents~\cite{Faessler:2006ky};

- We presented a detailed analysis of the light baryon observables
  all of which are in good agreement with experimental data.

\vspace*{.5cm}

This work was supported by the DFG under contracts FA67/31-1 and
GRK683. This research is also part of the EU Integrated
Infrastructure Initiative Hadronphysics project under contract
RII3-CT-2004-506078 and President grant of Russia
"Scientific Schools" No.5103.2006.2. K.P. thanks the Development
and Promotion of Science and Technology Talent Project (DPST),
Thailand for financial support. BRH is supported by the US
National Science Foundation under Grant No. PHY 02-44801 and would
like to thank the T\"ubingen theory group for its hospitality.

\end{document}